\documentclass{article}
\pdfoutput=1

\usepackage{amssymb,amsfonts,amsmath}
\usepackage{cite,enumerate,float}
\usepackage{color}
\usepackage{tikz}
\usetikzlibrary{arrows,snakes,backgrounds}

\usepackage{amssymb,amsfonts,amsmath,stmaryrd}
\usepackage{cite,enumerate,float,indentfirst}
\usepackage{color}
\usepackage{tikz}
\usetikzlibrary{arrows,snakes,backgrounds,calc}
\usepackage{ytableau}
\usepackage[vcentermath]{youngtab}

\def\be{\begin{eqnarray}}
\def\ee{\end{eqnarray}}
\def\nn{\nonumber}

\def\be{\begin{eqnarray}}
\def\ee{\end{eqnarray}}
\def\nn{\nonumber}

\def\p{\partial}
\def\tr{{\rm tr}\,}
\def\Tr{{\rm Tr}\,}

\def\B{W}
\def\c{\pi}

\definecolor{red}{rgb}{1,0,0}
\definecolor{orange}{rgb}{1,0.5,0}
\definecolor{violet}{rgb}{0.7,0,1}

\hoffset= - 1.0in         % switch off for draft style
\begin{document}
\title{\vspace{1.5cm}\bf
Summing up perturbation series around\\ superintegrable point
}

\author{
A. Mironov$^{b,c,d,}$\footnote{mironov@lpi.ru,mironov@itep.ru},
A. Morozov$^{a,c,d,}$\footnote{morozov@itep.ru},
A. Popolitov$^{a,c,d,}$\footnote{popolit@gmail.com},
Sh. Shakirov$^{d,}$\footnote{shakirov.work@gmail.com}
}

\date{ }

\maketitle

\vspace{-6.5cm}

\begin{center}
\hfill FIAN/TD-02/24\\
\hfill IITP/TH-01/24\\
\hfill ITEP/TH-01/24\\
\hfill MIPT/TH-01/24
\end{center}

\vspace{4.5cm}

\begin{center}
$^a$ {\small {\it MIPT, Dolgoprudny, 141701, Russia}}\\
$^b$ {\small {\it Lebedev Physics Institute, Moscow 119991, Russia}}\\
$^c$ {\small {\it NRC ``Kurchatov Institute", 123182, Moscow, Russia}}\\
$^d$ {\small {\it Institute for Information Transmission Problems, Moscow 127994, Russia}}
\end{center}

\vspace{.1cm}

\begin{abstract}
We work out explicit formulas for correlators in the Gaussian matrix model perturbed by
a logarithmic potential, i.e. by inserting Miwa variables.
In this paper, we concentrate on the example of a single Miwa variable.
The ordinary Gaussian model is superintegrable, i.e. the average of the Schur functions $S_Q$ is an
explicit function of the Young diagram $Q$.
The question is what happens to this property after perturbation.
We show that the entire perturbation series can be nicely
summed up into a kind of Borel transform of a universal exponential function,
while the dependence on $R$ enters through a polynomial factor
in front of  this exponential.
Moreover, these polynomials can be described explicitly through a single additional
structure, which we call ``truncation'' of the Young diagram $Q$.
It is unclear if one can call this an extended superintegrability,
but at least it is a tremendously simple deformation of it.
Moreover, the vanishing Gaussian correlators remain vanishing and, hence, are not deformed at all.
\end{abstract}

\bigskip

\section{Introduction}

Superintegrability in quantum field theory was originally defined in \cite{IMM,MM} (based on the
phenomenon earlier observed in \cite{DiF}--\cite{Pop}, see also some preliminary results in \cite{Kaz}--\cite{MKR} and later progress in \cite{MMten}--\cite{MO},\cite{MMssi}--\cite{CMPT})
as a possibility to find a basis in the space of correlators (typically, they are some characters) , when they can be
explicitly calculated.
The basis can be a somewhat transcendental (not expressed in elementary functions),
but just a little, unlike the case of generic non-superintegrable models.
This was supposed to mimic the situation in superintegrable classical potentials,
like the harmonic oscillator or Newton/Coulomb potentials, when the orbits become
periodic and are expressed in terms of periodic, though still elliptic integrals
(i.e. not just elementary trigonometric functions).
As a basic example in QFT, we took the Gaussian matrix model \cite{MM},
other examples include Selberg (logarithmic) models \cite{Sel}--\cite{Kad2},\cite{MMSh,MMShS} and a variety
of other theories \cite{MMsi}.
This basic example states that the Gaussian average of the Schur function $S_R$ is explicitly calculable
in terms of the same Schur functions:
\be\label{siG}
\Big< S_R \Big> = \eta_R(N) S_R\{\delta_{k,2}\}
\ee
with $\eta_R(N):= \frac{S_R[N]}{S_R\{\delta_{k,1}\}}$,
where the Schur function is labeled by the Young diagram $R$ and is a graded polynomial of time variables,
which is expressed at the l.h.s. through ``the quantum fields"  $p_k = \tr X^k$, $X$ being a
$N\times N$ Hermitian matrix in the case of Gaussian matrix model. The correlators in this latter are defined
\be
\Big<\ldots \Big> =  \int_{N\times N}  dX e^{-\frac{1}{2}\tr X^2} \ \ldots
\ee
and are normalized in such a way that $<1>=1$.
At the r.h.s. of the Schur function in (\ref{siG}), the variables are restricted to particular loci $p_k = \delta_{k,m}$ and $p_k = N$.
The ratio $\eta_R(N)$ is just a polynomial in $N$, a product over the boxes $(i,j)$ of the Young diagram:
$\eta_R = \prod_{(i,j)\in R} (N+i-j)$.

In this paper, we are going to consider a deformation of the Gaussian model\footnote{In the deformed case, we do not change the normalization of correlators as compared with the non-deformed case in order to have formulas simpler.}:
\be\label{dGm}
\Big<\ldots \Big>_{\c_k}=\int dX\exp\left(-{1\over 2}\Tr X^2+\sum_k{\c_k\over k}\Tr X^k\right) \ \ldots
\ee
The r.h.s. of this expression is understood as a power series in $\c_k$'s, i.e. this is an arbitrary deformation that is associated with the same integration contour along the real axis as in the Gaussian model. In other words, this is a small deformation over the Gaussian background, and parameters $\{\c_k\}$ are considered as describing small deviations from the Gaussian action. Here we consider only the specialization $\c_k=\pm z^k$ and explain the structure of correlators in such a model. It appears to have an interesting and intriguing feature: it turns out that the correlators are given by a Borel transform of a finite degree polynomial times a quadratic exponential.

Let us explain the origin of this Borel transform. To this end, we notice that, in the Gaussian model, the generating function of all correlators looks like
\be
Z\{p_k\} :=  \sum_R S_R\{p_k\} \Big< S_R \Big>
= \sum_R  \eta_R(N) S_R\{\delta_{k,2}\}S_R\{p_k\}
\ee
If not this $\eta_R(N)$, one would get an elementary answer, using the Cauchy formula
\cite{Mac}:
\be
\sum_R    S_R\{\bar p_k \}S_R\{p_k\} = \exp\left(\sum_k \frac{\bar p_k p_k}{k}\right) \ \
\Longrightarrow \ \
\sum_R    S_R\{\delta_{k,2}\}S_R\{p_k\} = e^{p_2/2}
\ee
As explained in \cite{China1,China2}, insertion of the polynomial factor $\eta_R(N)$ can be induced
by an action of a linear operator acting on the time variables $p_k$:
\be
Z\{p_k\} = \widehat{\cal O}(N)\left\{ \sum_R    S_R\{\delta_{k,2}\}S_R\{p_k\}\right\} =
\widehat{\cal O}(N)  e^{p_2/2}
\ee
The action of operator $\widehat{\cal O}$ can be interpreted as an {\it enhanced} Borel transform.
Usually Borel ``improves" infinite series by inserting extra factorials \cite{Borel}
or their combinatorial counterparts \cite{MMBorel} in denominators.
In the case of polynomial insertions like $\eta_R$,
the series are sometimes\footnote{In the case of choosing the deformation of the Gaussian model by $\c_k=-\sum_az_a^k$, see sec.5 below.} {\it cut off} by a combination of $\Gamma$-function factors $\frac{N!}{(N-n)!}$
at finite values of $n$, actually regulated by the size $N$ of the matrix in
the underlying matrix model.
This cutoff is a kind of extreme (enhanced) version of the same convergency-improvement idea.

The paper is organized as follows.
In section 2, we discuss a structure of correlators in the Gaussian model with general deformation,
and then which form it takes if the deformation is parameterized by Miwa variables.
In section 3, we consider in detail the correlators $\Big< S_Q\Big>_{\c_k=z^k}$
in the model deformed by a {\it single} Miwa variable,
and describe an interesting structure of this perturbation expansion.
Coefficients are expressed in terms of peculiar polynomials $\B_j({\bf Q})$
depending on somewhat mysterious truncation $Q\longrightarrow {\bf Q}$ of the Young diagram $Q$.
Summation over $j$ involves factorials which can be treated as application of an ``enhanced Borel transform",
which we define and discuss in section 4. In section 5, we consider the correlators $\Big< S_Q\Big>_{\c_k=-z^k}$.
A brief conclusion in section 6 sets the problems for further studies in this  direction.
The Appendix contains some evidence in support of the representation of correlators in terms of $\B_j({\bf Q})$ in section \ref{secMiwa1}.

\paragraph{Notation.} We use the notation $S_R\{p\}$ for the Schur functions, which are graded polynomials of arbitrarily many variables $p_k$ of grading $k$. The Schur function $S_R\{p\}$ is labelled by the Young diagram (partition) $R$: $R_1\ge R_2\ge\ldots\ge R_{l_R}>0$, and the grading of this Schur function is $|R|:=\sum_iR_i$. Similarly, we denote $S_{R/P}\{p\}$ the skew Schur functions \cite{Mac}.

Throughout the paper, we use the Pocchammer symbol
\be\label{Poch}
(N;\mu)_n=\prod_{k=0}^{n-1} (N+k\mu)
\ee
and $(N;\mu)_0=1$.

\section{Miwa deformation of superintegrability for Gaussian model}

\subsection{Deformation of Gaussian model}

Thus, in this paper, we consider the correlators in the deformed Gaussian model. The generation function of the correlators is given by
\be
Z=\int dX\exp\left(-{1\over 2}\Tr X^2+\sum_k{p_k+\c_k\over k}\Tr X^k\right)
\label{p+c}
\ee
and, using (\ref{siG}), we immediately obtain
\be\label{Z}
Z=\sum_R\eta_R(N)S_R\{\delta_{k,2}\}S_R\{p_k+\c_k\}=\sum_{R,Q}\eta_R(N)S_R\{\delta_{k,2}\}S_{R/Q}\{\c_k\}S_Q\{p_k\}
\ee
i.e., for an arbitrary correlator, one obtains
\be\label{gc}
\left<S_Q\{\Tr X^k\}\right>_{\c_k}=\sum_R\eta_R(N)S_R\{\delta_{k,2}\}S_{R/Q}\{\c_k\}
\ee
This is an infinite sum. It is always divisible by to $\eta_Q$, since the summand is non-zero only if the Young diagram $R$ contains all boxes of the Young diagram $Q$ inside. Hence, the correlator can be written as
\be
\left<S_Q\{\Tr X^k\}\right>_{\c_k}=\eta_Q(N)F_Q(\c_k,N)
\ee
where
\be\label{FQ}
F_Q(\c_k,N):=\sum_R\frac{\eta_R(N)}{\eta_Q(N)}S_R\{\delta_{k,2}\}S_{R/Q}\{\c_k\}
\label{FQdef}
\ee
is a power series in $N$ and $c_k$.

For $\c_k=0$ and $F_Q=S_R\{\delta_{k,2}\}$, one obtains the usual superintegrability formula (\ref{siG}) \cite{MMsi},
for $\c_k\neq 0$, one gets a {\it deformation}, which we can treat as that of superintegrability
-- provided $F_Q(\c_k,N)$ are calculable and simple enough.

\subsection{Gaussian model deformed by Miwa variables}

As we already explained, the Miwa parametrization of the deforming constants $\c_k=\sum_iz_i^k$ looks quite natural. In this case, one
inserts in the Gaussian integral and additional factor of
\be
\prod_{i=1}^m{1\over \det(1-z_iX)}
\ee
In the case of generic $m$, formula (\ref{FQ}) for $F_Q(\c_k,N)$ does not look simple (though it possesses an interesting structure that we will discuss elsewhere), however, in the case of $m=1$, it does:
\be
F_Q(z,N)=\sum_Rz^{|R|-|Q|}\frac{\eta_R(N)}{\eta_Q(N)}S_R\{\delta_{k,2}\}S_{R/Q}\{1\}
\ee
where $|R|=\sum_iR_i$ denotes the size of the Young diagram $R$.

First few examples of the function $F_Q(z,N)$ are:
\be
F_{[k+1]}(z,N)={1-(-1)^k\over 2}{k!!\over (k+1)!}+{1\over k!}\sum_{i=[{k+1\over 2}]}(2i-1)!!z^{2i+1-k} +O(N)
\ee
where $[...]$ denotes the integer part of a number.
Since deformation with the Miwa variable corresponds to inserting $\det(1-zX)^{-1}$ into the matrix integral measure, the integral diverges, hence, the sum also diverges. One of the possibilities to deal with it is to notice that inserting the regular expression $\det(1-zX)$ corresponds just to $c_k=-z^k$, and to use that $S_{R/Q}\{-c_k\}=(-1)^{|R|+|Q|}S_{R^\vee/Q^\vee}\{c_k\}$, where $R^\vee$ denotes the conjugated Young diagram. Then,
\be
F_{[1^{k+1}]}(z,N)={1-(-1)^k\over 2}{(-1)^{k+1\over 2}k!!\over (k+1)!}+{1\over k!}\sum_{i=[{k+1\over 2}]}(-1)^{i+1}(2i-1)!!z^{2i+1-k} +O(N)
\ee
Another possibility is to introduce the Miwa variable with an arbitrary multiplicity $\mu$: $c_k=\mu z^k$ so that, for instance,
\be
F_{[1]}(z,N)=\sum_{n=1}z^n\sum_{g=0}^{[{n\over 2}]}C(n,g)\mu^{2g+1+2\{{n+1\over 2}\}} z^{2n-1} +O(N)
\ee
where $\{...\}$ denotes the fractional part of a number, and the numbers $C(k,n)$ are given by the sequence A035309 \cite{oeis,AkhSh}:
\be
C(n,g)={(2n)!\over (n+1)!(n-2g)!}\beta_{2g}^{(n)}\nn\\
\left({x\over 2\tanh{x\over 2}}\right)^{n+1}=\sum_{k=0}\beta_k^{(n)}x^k
\ee

\section{Gaussian model deformed by one Miwa variable
\label{secMiwa1}}

In this section, we present a systematic description of the correlators in the case of the Gaussian model deformed by one Miwa variable.
It turns out that one can evaluate the infinite sums, (\ref{FQ}) for $F_Q(z,N)$ in this case.

\subsection{Rectangular correlators with one Miwa deformation}

We start with correlators described by rectangular Young diagrams. The simplest ones are
\be
F_{[2r]} = \sum_{k=0}^\infty \sum_{j=0}^{{\rm min}(k,r)} \frac{(N+2j,1)_{2k-2j}z^{2k}}{(2r-2j)!!(2k-2j)!!}
= \sum_{k=0}^\infty  \sum_{j=0}^{{\rm min}(k,r)} \frac{(N+2k-1)!}{(N+2j-1)!}\frac{z^{2k}}{(2r-2j)!!(2k-2j)!! } \nn \\
F_{[2r+1]} = \sum_{k=0}^\infty \sum_{j=0}^{{\rm min}(k,r)} \frac{(N+2j+1,1)_{2k-2j}z^{2k+1}}{(2r-2j)!!(2k-2j)!!}
= \sum_{k=0}^\infty  \sum_{j=0}^{{\rm min}(k,r)} \frac{(N+2k)!}{(N+2j)!} \frac{z^{2k+1}}{(2r-2j)!!(2k-2j)!!}
\ee
and
\be\label{13}
F_{[1^{2s}]} = \frac{(-)^s}{(2s)!!} \sum_{k=0}^\infty  \frac{(N+2k-1)!}{(N-1)!}\frac{z^{2k}}{(2k)!!} \nn \\
F_{[1^{2s+1}]} = \frac{(-)^s}{(2s)!!} \sum_{k=0}^\infty  \frac{(N+2k)!}{N!}\frac{z^{2k+1}}{(2k)!!}
\ee
Further examples reveal the general structure of the correlators:
\be
F_{[2^{2s}]} = \frac{1}{[(2s)!!]^2 } \sum_{k=0}^\infty
\left(\frac{(N+2k-1)!}{(N-1)!}+ 4sk\frac{(N+2k-1)!}{(N+1)!}\right)\frac{z^{2k}}{(2k)!!} \nn \\
F_{[2^{2s+1}]} = \frac{1}{(2s)!!(2s+2)!!} \sum_{k=0}^\infty
\left(\frac{(N+2k-1)!}{(N-1)!}+ 4(s+1)k\frac{(N+2k-1)!}{(N+1)!}\right)\frac{z^{2k}}{(2k)!!}
\ee
\be
 F_{[3^{2s}]} = \frac{2(-)^s}{[(2s)!!]^2 (2s+2)!! } \sum_{k=0}^\infty
\left(\frac{(N+2k-1)!}{(N-1)!}+ 4sk\frac{(N+2k-1)!}{(N+1)!}\right)\frac{z^{2k}}{(2k)!!} \nn \\
  F_{[3^{2s+1}]} = \frac{2(-)^s}{(2s)!![(2s+2)!!]^2} \sum_{k=0}^\infty
\left(\frac{(N+2k)!}{N!}+ 4(s+1)k\frac{(N+2k)!}{(N+2)!}\right)\frac{z^{2k+1}}{(2k)!!}
\ee
{\footnotesize
\be
 F_{[4^{2s}]} = \frac{4 }{[(2s)!!]^2 [(2s+2)!!]^2  } \sum_{k=0}^\infty
\left(\frac{(N+2k-1)!}{(N-1)!}+  8sk\frac{(N+2k-1)!}{(N+1)!}
+  16s(s+1)k(k-1)\frac{(N+2k-1)!}{(N+3)!}\right)\frac{z^{2k}}{(2k)!!} \nn \\
\!\!\!\!\!\!\!\!  F_{[4^{2s+1}]} = \frac{4}{(2s)!![(2s+2)!!]^2(2s+4)!!} \sum_{k=0}^\infty
\left(\frac{(N+2k-1)!}{(N-1)!}+  8(s+1)k\frac{(N+2k-1)!}{(N+1)!}
+ 16(s+1)(s+2)k(k-1)\frac{(N+2k-1)!}{(N+3)!}\right)\frac{z^{2k}}{(2k)!!}
\nn
\ee
}
Generic expression for rectangular diagrams is
\be\label{cr2}
F_{[(2r)^{2s}]} = \Big(\Xi_0^r\Big)^2 \sum_{k=0}^\infty  (N;1)_{2k}
\sum_{j=0}4^j\binom{r}{j}{(s;1)_j(k;-1)_j\over (N;1)_{2j}}\frac{z^{2k}}{(2k)!!}\nn\\
F_{[(2r)^{2s+1}]} = \Xi_0^r\Xi_1^{r+1}
\sum_{k=0}^\infty  (N;1)_{2k}\sum_{j=0}4^j\binom{r}{j}{(s+1;1)_j(k;-1)_j\over (N;1)_{2j}}\frac{z^{2k}}{(2k)!!}\nn\\
F_{[(2r+1)^{2s}]} =(-1)^s \Xi_0^r\Xi_0^{r+1}\sum_{k=0}^\infty  (N;1)_{2k}
\sum_{j=0}4^j\binom{r}{j}{(s;1)_j(k;-1)_j\over (N;1)_{2j}}\frac{z^{2k}}{(2k)!!}\nn\\
F_{[(2r+1)^{2s+1}]} = {(-1)^s \over (2r)!!}\Xi_0^{r+1}\Xi_1^{r+1} \sum_{k=0}^\infty  (N+1;1)_{2k}
\sum_{j=0}4^j\binom{r}{j}{(s+1;1)_j(k;-1)_j\over (N+1;1)_{2j}}\frac{z^{2k+1}}{(2k)!!}
\ee
where
\be
\Xi_a^r=\prod_{j=a}^{r-1}\left({(2j)!!\over (2s+2j)!!}\right)
\ee
It follows from (\ref{FQ}) that the common coefficients in the first three formulas are just $S_R\{\delta_{k,2}\}$:
\be
S_{[(2r)^{2s}]}\{\delta_{k,2}\} = \Big(\Xi_0^r\Big)^2\nn\\
S_{[(2r)^{2s+1}]}\{\delta_{k,2}\} = \Xi_0^r\Xi_1^{r+1}\nn\\
S_{[(2r+1)^{2s}]}\{\delta_{k,2}\} =(-1)^s \Xi_0^r\Xi_0^{r+1}
\ee
As to the fourth coefficient, it is equal to
\be
2(r+s+1)S_{[2r+2,(2r+1)^{2s}]}\{\delta_{k,2}\}={(-1)^s \over (2r)!!}\Xi_0^{r+1}\Xi_1^{r+1}
\ee
Indeed, only two terms contribute to formula (\ref{FQ}) in this case, those with $R$'s that contain the diagram $Q=[(2r+1)^{2s+1}]$ and have one box more:
\be
(N+2r+1)S_{[2r+2,(2r+1)^{2s}]}\{\delta_{k,2}\} +(N-2s-1)S_{[(2r+1)^{2s+1},1]}\{\delta_{k,2}\}
={(-1)^s \over (2r)!!}\Xi_0^{r+1}\Xi_1^{r+1}
\ee
where we used that\footnote{The first equality is evident at $r=s$ because of the transposition rule for the Schur functions \cite{Mac}.}
\be
S_{[2r+2,(2r+1)^{2s}]}\{\delta_{k,2}\}=-S_{[(2r+1)^{2s+1},1]}\{\delta_{k,2}\}={(-1)^s \over 2(r+s+1)(2r)!!}\Xi_0^{r+1}\Xi_1^{r+1}
\ee
This kind of formulas are derived using (\ref{S2}).

\subsection{
Generic correlators
}

To provide some impression of what the non-rectangular formulas look like,
we begin from a simple example:
\be
F_{[2,1^{2s+1}]} = 0 \nn \\
F_{[2 ]} =  \frac{1}{2}
\sum_{k=0}^\infty \left( \frac{(N+2k-1)!}{(N-1)!} + 4k \frac{(N+2k-1)!}{(N+1)!}\right)
\frac{z^{2k}}{(2k)!!} \nn \\
F_{[2,1^{2s-2}]} =  \frac{(-1)^{s+1}}{(2s)!!}
\sum_{k=0}^\infty \left( \frac{(N+2k-1)!}{(N-1)!} + 4sk \frac{(N+2k-1)!}{(N+1)!}\right)
\frac{z^{2k}}{(2k)!!}
\ee
The new phenomenon here is that some $F_Q$ are identically vanishing.
The structure of other, however, does not really deviate from what we already found
for rectangular $Q$.

\bigskip

The answer for $F_Q$ of even size $|Q|$ is
\be\label{FQeven}
\boxed{
F_Q = S_Q\{\delta_{k,2}\}\cdot
\sum_{k=0}^\infty \frac{z^{2k}}{(2k)!!} \sum_{j=0}^k c_{j}(Q) \frac{k!}{(k-j)!}\frac{(N+2k-1)!}{(N+2j-1)!}=
S_Q\{\delta_{k,2}\}\cdot
\sum_{k=0}^\infty (N;1)_{2k}\frac{z^{2k}}{(2k)!!} \sum_{j=0}^k c_{j}(Q) \frac{(k;-1)_j}{(N;1)_{2j}}
}
\ee
while that for $F_Q$ of odd size $|Q|$ is
\be
\boxed{
F_Q =\underbrace{S'_Q\{\delta_{k,2}\}}_{C_Q}\cdot
\sum_{k=0}^\infty \frac{z^{2k+1}}{(2k)!!} \sum_{j=0}^k c_{j}(Q) \frac{k!}{(k-j)!}\frac{(N+2k)!}{(N+2j)!}=
S'_Q\{\delta_{k,2}\}\cdot
\sum_{k=0}^\infty (N;1)_{2k+1}\frac{z^{2k+1}}{(2k)!!} \sum_{j=0}^k c_{j}(Q) \frac{(k;-1)_j}{(N;1)_{2j+1}}
}
\label{FQodd}
\ee

Now we explain various ingredients of these formulas.

\paragraph{Overall coefficients.}
The overall coefficient in the case of even $|Q|$ in (\ref{FQeven}) follows immediately from (\ref{FQ}): the coefficient in front of $z^0$ is determined by $R=Q$ and thus is equal just to $S_Q\{\delta_{k,2}\}$.

The overall coefficient $C_Q$ in the case of odd $|Q|$ is proportional to $z$, i.e. only the Young diagrams $R$ with $|R|=|Q|+1$ contributes to the sum (\ref{FQ}):
\be
C_Q=\sum_{Q+\Box}(N+j_\Box-i_\Box)S_{Q+\Box}\{\delta_{k,2}\}
\ee
Note that this coefficient does not depend on $N$. Indeed, because of the Pieri rule \cite{Mac}
\be
p_1S_Q\{p_k\}=\sum_{Q+\Box}S_{Q+\Box}\{p_k\}
\ee
we immediately obtain
\be
\sum_{Q+\Box}S_{Q+\Box}\{\delta_{k,2}\}=0
\ee
Hence,
\be
C_Q=\sum_{Q+\Box}(j_\Box-i_\Box)S_{Q+\Box}\{\delta_{k,2}\}
\ee
Note that this sum can be reproduced by the action of the differential operator
$\sum_kkp_{k+1}{\p\over\p p_k}$, \cite{China1}:
\be
C_Q=\sum_{Q+\Box}(j_\Box-i_\Box)S_{Q+\Box}\{\delta_{k,2}\}=\sum_kkp_{k+1}{\p\over\p p_k}S_Q\{p_k\}\Big|_{p_k=\delta_{k,2}}={\p\over\p p_1}S_Q\{p_k\}\Big|_{p_k=\delta_{k,2}}
:= S'_Q\{\delta_{k,2}\}
\label{CviaSprime}
\ee
We substituted this answer into (\ref{FQodd}), so that the only remaining question about  $F_Q$
concerns the coefficients $c_j(Q)$, which we will now describe.

\paragraph{Coefficients $c_j(Q)$.} The coefficients $c_j(Q)$ are expressed through auxiliary functions
\be
\B^{(1)}_j(a) = 4^j \frac{a!}{(a-j)!} = 4^j \prod_{i=0}^{j-1} (a-i) \nn \\
 \B^{(2)}_j(a,b):=4^j \sum_{k=0}^j \left(\prod_{i=0}^{k-1} (b-i) \prod_{i=k}^{j-1} (a-i)\right) \nn \\
 \B^{(3)}_j(a,b,c):=4^j \sum_{0\leq k_1\leq k_2\leq j}
 \left(\prod_{i=0}^{k_1-1} (c-i)\prod_{i=k_1}^{k_2-1} (b-i) \prod_{i=k_2}^{j-1} (a-i)\right) \nn \\
 \ldots
\label{Wpolsexa}
\ee
and generically
\be
\B_j^{(m)}(a_1\geq a_2 \geq \ldots \geq a_m) :=\ 4^j\cdot\!\!\!\!\!\!\!\!\!\!\!\!\!\!
\sum_{0\leq k_1\leq\ldots\leq k_{m-1}\leq j}\left( \prod_{i=0}^{k_1-1} (a_m-i)  \prod_{i=k_1}^{k_2-1} (a_{m-1}-i) \ldots
\prod_{i = k_{m-1}}^{j-1} (a_1-1)\right)=\nn\\
=4^j(a_1;-1)_j\sum_{j\ge k_1\ge k_2\ldots\ge k_{m-1}\ge 0}\prod_{i=1}^{m-1}{(a_{i+1};-1)_{k_i}\over(a_i;-1)_{k_i}}
\label{Wpols}
\ee
where
\be
V_j(a,b)=\sum_{k=0}^j{(b;-1)_k\over(a;-1)_k}
\ee
Thus,
\be
\B^{(m)}_j(a_1,\ldots,a_{m-1},a_m)=\B^{(m-1)}_j(a_1,\ldots,a_{m-1})V_j(a_{m-1},a_m)
\ee
and, since $V_j(a,0)=1$, these polynomials are connected by a simple reduction:
\be\label{trQ}
\B_j^{(m)}(a_1\geq a_2 \geq \ldots \geq a_{m-1}\geq 0) = \B_j^{(m-1)}(a_1\geq a_2 \geq \ldots \geq a_{m-1})
\ee

Explicit expressions for diagrams of the odd size $|Q|$ are:
\be\label{mpodd}
\text{\bf 1-point function}
&c_j([2a+1])
= \B^{(1)}_j(a)  \nn \\
\nn \\
\text{\bf 2-point functions}
& c_j([2a+1,2b])
= \B^{(2)}_j(a,b)\nn \\
&F_{[2a,2b+1]} = 0 \nn \\
\nn \\
\text{\bf 3-point functions}
& c_j([2a+1,2b+1,2c+1]) = \B^{(2)}_j(a,c)  \nn \\
& c_j([2a+1,2b,2c]) = \B^{(2)}_j(a,b)     \nn \\
& c_j([2a,2b,2c+1]) = \B^{(2)}_j(b-1,c)  \nn \\
& F_{[2a,2b+1,2c]} = 0   \nn \\
\nn \\
\text{\bf 4-point functions}
&c_j([2a+1,2b,2c,2d]) = \B^{(3)}_j(a,b,d)   \nn \\
&c_j([2a,2b,2c+1,2d]) = \B^{(3)}_j(b-1,c,d)   \nn \\
&c_j([2a+1,2b+1,2c+1,2d]) = \B^{(3)}_j(a,c,d)   \nn \\
&c_j([2a+1,2b,2c+1,2d+1]) = \B^{(3)}_j(a,b,c+1)   \nn \\
&F_{[2a,2b+1,2c,2d]} = 0  \nn \\
&F_{[2a,2b,2c,2d+1]}= 0   \nn \\
&F_{[2a+1,2b+1,2c,2d+1]} = 0\nn \\
&F_{[2a,2b+1,2c+1,2d+1]} =0  \nn \\
\ldots
\ee
and those for the even size $|Q|$ are:
\be\label{mpeven}
\text{\bf 1-point function}
&c_j([2a])
= \B^{(1)}_j(a) \nn \\
\nn \\
\text{\bf 2-point functions}
&c_j([2a,2b])
= \B^{(1)}_j(a) \nn \\
&c_j([2a+1,2b+1])
= \B^{(1)}_j(b) \nn \\
\nn \\
\text{\bf 3-point functions}
&c_j([2a,2b,2c]) =
\B^{(2)}_j(a,c) \nn \\
&c_j([2a+1,2b+1,2c]) =  \B^{(2)}_j(a,c) \nn\\
&c_j([2a,2b+1,2c+1]) =  \B^{(2)}_j(a,b+1)   \nn \\
&F_{[2a+1,2b,2c+1]} = 0  \nn \\
\nn \\
\text{\bf 4-point functions}
&c_j([2a,2b,2c,2d]) =\B^{(2)}_j(a,c) \nn \\
&c_j([2a+1,2b+1,2c+1,2d+1]) =\B^{(2)}_j(b,d)\nn\\
&c_j([2a+1,2b+1,2c,2d]) =\B^{(2)}_j(a,c) \nn \\
&c_j([2a+1,2b,2c,2d+1]) =\B^{(2)}_j(c-1,d) \nn \\
&c_j([2a,2b+1,2c+1,2d]) =\B^{(2)}_j(a,b+1) \nn \\
&c_j([2a,2b,2c+1,2d+1]) =\B^{(2)}_j(a,d) \nn \\
&c_j([2a+1,2b,2c+1,2d]) =0\nn \\
&F_{[2a,2b+1,2c,2d+1])}=0 \nn \\
\ldots
\ee
With some abuse of terminology we ascribe to  the averages $F_Q$ for $Q$ with $m$-lines the name ``$m$-point functions'',
still it looks natural in this context.
There are a lot of vanishing $F_Q$'s because of vanishing the overall coefficient when it makes no sense to discuss $c_j(Q)$, these cases are manifestly indicated. In fact,
\be
F_{[2a_1+1,2a_2+1,2a_3,2a_4+1,2a_5,\ldots]} =0\nn\\
F_{[2a_1+1,2a_2,2a_3+1,2a_4,\ldots]} =0\nn\\
F_{[2a_1,2a_2,\ldots,2a_{2s+1},2a_{2s}+1,\ldots]} =0\ \ \ \ \ \ \forall\ s,\ \ \ \ \hbox{odd }|Q| \nn\\
\ldots
\ee
where $\ldots$ denote lines of arbitrary parity.

Generally, if the number of lines in the Young diagram at even $|Q|$ is $2s$, then $c_j(Q)=W_j^{(s)}$, and if it is $2s+1$, then $c_j(Q)=W_j^{(s+1)}$. At the same time, at odd $|Q|$ it is $c_j(Q)=W_j^{(s+1)}$ in the both cases.

In the case of all even  lines in the Young diagram $Q$, the answer looks like
\be
c_j([2a_1,2a_2,\ldots,2a_{2s}]) =\B^{(s)}_j(a_1,a_3,\ldots,a_{2s-1})\nn\\
c_j([2a_1,2a_2,\ldots,2a_{2s+1}]) =\B^{(s+1)}_j(a_1,a_3,\ldots,a_{2s+1})
\ee
Likewise for all odd lines
\be
c_j([2a_1+1,2a_2+1,\ldots,2a_{2s}+1]) =\B^{(s)}_j(a_2,a_4,\ldots,a_{2s})\nn\\
c_j([2a_1+1,2a_2+1,\ldots,2a_{2s+1}+1]) =\B^{(s+1)}_j(a_1,a_3,\ldots,a_{2s+1})
\ee
One can immediately generate more general series:
\be
c_j([2a_1+1,2a_2,2a_3,\ldots,2a_{2s}]) =\B^{(s+1)}_j(a_1,a_2,a_4,a_6\ldots,a_{2s})\nn\\
c_j([2a_1+1,2a_2,2a_3,\ldots,2a_{2s+1}]) =\B^{(s+1)}_j(a_1,a_2,a_4,\ldots,a_{2s})\nn\\
c_j([2a_1+1,2a_2+1,2a_3,2a_4,\ldots,2a_{2s}]) =\B^{(s)}_j(a_1,a_3,\ldots,a_{2s-1})\nn\\
c_j([2a_1+1,2a_2+1,2a_3,2a_4,\ldots,2a_{2s+1}]) =\B^{(s+1)}_j(a_1,a_3,\ldots,a_{2s+1})\nn\\
c_j([2a_1,2a_2+1,2a_3+1,2a_4,2a_5,\ldots,2a_{2s}]) =\B^{(s)}_j(a_1,a_2+1,a_5,a_7,\ldots,a_{2s-1})\nn\\
c_j([2a_1,2a_2+1,2a_3+1,2a_4,2a_5,\ldots,2a_{2s+1}]) =\B^{(s+1)}_j(a_1,a_2+1,a_5,a_7,\ldots,a_{2s+1})\nn\\
c_j([2a_1+1,2a_2+1,2a_3+1,2a_4,2a_5,\ldots,2a_{2s}]) =\B^{(s+1)}_j(a_1,a_3,a_4,a_6,\ldots,a_{2s})\nn\\
c_j([2a_1+1,2a_2+1,2a_3+1,2a_4,2a_5,\ldots,2a_{2s+1}]) =\B^{(s+1)}_j(a_1,a_3,a_4,a_6,\ldots,a_{2s})\nn\\
c_j([2a_1,2a_2,2a_3+1,2a_4,2a_5,\ldots,2a_{2s}]) = \B^{(s+1)}_j(a_2-1,a_3,a_4,a_6,\ldots,a_{2s})\nn\\
c_j([2a_1,2a_2,2a_3+1,2a_4,2a_5,\ldots,2a_{2s+1}]) = \B^{(s+1)}_j(a_2-1,a_3,a_4,a_6,\ldots,a_{2s})
\ee
etc.

Note that formulas (\ref{cr2}) correspond to
\be
c_j[(2r)^{2s}]=\B_j^{(s)}(\underbrace{r,r,\ldots,r}_{s\ times})\nn\\
c_j[(2r)^{2s+1}]=\B_j^{(s+1)}(\underbrace{r,r,\ldots,r}_{s+1\ times})\nn\\
c_j[(2r+1)^{2s}]=\B_j^{(s)}(\underbrace{r,r,\ldots,r}_{s\ times})\nn\\
c_j[(2r+1)^{2s+1}]=\B_j^{(s+1)}(\underbrace{r,r,\ldots,r}_{s+1\ times})
\ee
It is also simple to understand when $c_j(Q)=0$: in the case of even $|Q|$, the overall coefficient is $S_Q\{\delta_{k,2}\}$.
The value of $S_Q\{\delta_{k,2}\}$ is equal \cite{Pop,MMNO} to
\be\label{S2}
S_Q\{\delta_{k,2}\}=\delta_2(Q)\prod_{(i,j)\in Q}{1\over h_{i,j}^{ev}}
\ee
where $h_{i,j}$ is the hook length, and the product runs over only hook with even length, which we denoted by the superscript $ev$. $\delta_2(Q)$ is defined in \cite[Eq.(3.26)]{Pop}:
\be
\delta_2(Q)=\left\{
\begin{array}{cl}
(-1)^{|Q|/2}\prod_{(i,j)\in Q}(-1)^{[c_{i,j}/2]+[h_{i,j}/2]}&\hbox{ if the $2$-core of $Q$ is trivial}\cr\cr
0&\hbox{otherwise}
\end{array}
\right.
\ee
where $c_{i,j}$ is the content of the box $(i,j)$ in $Q$. Thus, $c_j(Q)=0$ if the $2$-core of $Q$ is non-trivial.

For more details behind these formulas see the Appendix.

\subsection{Truncation map}

Note that, because of (\ref{trQ}), the argument of $\B_j$ is actually a Young diagram, which we denote ${\bf Q}$.
Hence, the construction of coefficients obey the simple rule
\be
\boxed{
c_j(Q) = \B_j({\bf Q})
}
\label{trunc}
\ee
where ${\bf Q}$ is a certain {\bf truncation} of the diagram $Q$.
In other words, instead of (\ref{FQeven}) and (\ref{FQodd}) we can write
\be\label{FQeven1}
\text{even} \ |Q|: \ \ \ &
F_Q = S_Q\{\delta_{k,2}\}\cdot
\sum_{k=0}^\infty \frac{z^{2k}}{(2k)!!} \sum_{j=0}^k  \frac{k!}{(k-j)!}\frac{(N+2k-1)!}{(N+2j-1)!}\,\B_{j}({\bf Q})
\nn \\
\text{odd} \  |Q|: \ \ \ &
F_Q =S'_Q\{\delta_{k,2}\}\cdot
\sum_{k=0}^\infty \frac{z^{2k+1}}{(2k)!!} \sum_{j=0}^k \frac{k!}{(k-j)!}\frac{(N+2k)!}{(N+2j)!} \, \B_{j}({\bf Q})
\label{FQodd1}
\ee
We summarize the truncation map $Q\longrightarrow {\bf Q}$ in the following table:
\be
\begin{array}{cc|ccc|cc}
&&&&&\\
& \text{odd}\ |Q|&&&& \text{even}\ |Q| \\
&&&&& \\
\hline
&&&&& \\
& Q  &\longrightarrow &{\bf Q}&\longleftarrow &  Q \\
&&&&&\\
\hline
& [2a+1] &\longrightarrow  & [a] &\longleftarrow & [\underline{2a}] \\
\hline
&        && [a] &\longleftarrow & [\underline{2a},\underline{2b}] \\
&        && [a] &\longleftarrow & [2a+1,2b+1] \\
&&&&& \\
& [\underline{2a},2b+1]&\longrightarrow & \emptyset && \\
&&&&& \\
& [2a+1,\underline{2b}] &\longrightarrow & [a,b] && \\
\hline
& [2a+1,\underline{2b},\underline{2c}] &\longrightarrow &  [a,b] &\longleftarrow & [2a+1,2b+1,\underline{2c}] \\
&&&&&\\
&[2a+1,\underline{2b},2c+1] &\longrightarrow & \emptyset &&  \\
&[\underline{2a},2b+1,\underline{2c}] &\longrightarrow & \emptyset && \\
&&&&&\\
& [2a+1,2b+1,2c+1] &\longrightarrow & [a,c] &\longleftarrow & [\underline{2a},\underline{2b},\underline{2c}] \\
&&& [b-1,c] &\longleftarrow & [\underline{2a},\underline{2b},2c+1] \\
&&&[a,b+1] &\longleftarrow & [\underline{2a},2b+1,2c+1]\\
\hline
&&&&&\\
& && \ldots && \\
&&&&&\\
\hline
&&&&&\\
&&& [r^s] &\longleftarrow &  [\underline{(2r)}^{2s}] & \\
&&& [r^s] &\longleftarrow &  [(2r+1)^{2s}] & \\
&[(2r+1)^{2s+1}] &\longrightarrow & [r^{s+1}]  &\longleftarrow & [\underline{(2r)}^{2s+1}] & \\
&&&&& \\
\hline
&&&&&\\
&&&[a_1,a_3,\ldots, a_{2s-1}]&\longleftarrow &  [2a_{1}+1,2a_2+1, \ldots, 2a_{2s}+1] & \\
&&& [a_1,a_3,\ldots, a_{2s-1}] &\longleftarrow &  [\underline{2a_1},\underline{2a_2},\ldots,\underline{2a_{2s}}] & \\
&[2a_1+1,\ldots,   2a_{2s+1}+1] & \longrightarrow & [a_1,a_3,\ldots, a_{2s-1},a_{2s+1}] &\longleftarrow &
[\underline{2a_1},\underline{2a_2},\ldots, \underline{2a_{2s+1}}] & \\
&&&&&\\
\hline
&&&&&\\
& && \ldots && \\
%&&&&&\\
\label{trunctable}
\end{array}
\ee

\noindent
Even entries of $Q$ are underlined for convenience.

\bigskip

Evaluating $F_Q$ with the help of explicit functions $\B_j$ is very simple and fast,
while that via the Schur functions takes a lot of computer time.
Thus these {\bf expressions (\ref{trunc}) are} indeed {\bf very practical},
 once one knows the truncation map,
like in the cases enumerated in Table (\ref{trunctable}).

In order to emphasize this, we list a few first $j$ the
formulas (\ref{trunc}) for the coefficients $c_j(Q)$ through the truncated Young diagrams
${\bf Q}=[{\bf q}_1 \geq {\bf q}_2 \geq \ldots \geq {\bf q}_{max} >0]$.
In an another explicit form, expressions (\ref{Wpols}) for $\B_j({\bf Q})$ look as follows:
\be
c_0(Q) = 1 \nn \\
c_1(Q) = 4\cdot |{\bf Q}| = 4\cdot \sum_i {\bf q}_i \nn \\
c_2(Q) = 16\cdot \sum_i {\bf q}_i ({\bf q}_1+\ldots+{\bf q}_i -i) \nn \\
c_3(Q) = 64 \cdot \Big({\bf q}_1({\bf q}_1-1)({\bf q}_1-2)
+ {\bf q}_2({\bf q}_1^2+{\bf q}_1{\bf q}_2+{\bf q}_2^2 -4{\bf q}_1- 5{\bf q}_2+6)
+ \ldots\Big) = \nn \\
= 64\cdot\sum_i {\bf q}_i \cdot \left(\sum_{1\leq j\leq k\leq i} {\bf q}_j{\bf q}_k
- \sum_{1\leq j\leq i}(i+j+1){\bf q}_j + i(i+1) \right) \nn \\
c_4(Q) = 256\cdot\Big({\bf q}_1({\bf q}_1-1)({\bf q}_1-2)({\bf q}_1-3) + \ldots\Big)
%\nn\\
%\ldots
\label{cviaq}
\ee

\bigskip

A more sophisticated problem is to find $W_j^{(s)}$ once one has a set of $c_j(Q)$ for a concrete $Q$, 
i.e. to solve these non-linear equations w.r.t. ${\bf q}_i$ in order to find ${\bf Q}$ from a given set of $c_j(Q)$.

For example, one can observe that $c_j(Q)\neq 0$ for all $j\leq {\bf q}_1$,
i.e. the first entry ${\bf q}_1$ of ${\bf Q}$ is the number of non-vanishing $c_j(Q)$   minus one.
It can be conveniently formulated in terms of the polynomial $\sigma_Q(w) = \sum_k c_j(Q)w^j$,
namely,  ${\bf q}_1 = {\rm degree}_w(\sigma)$.

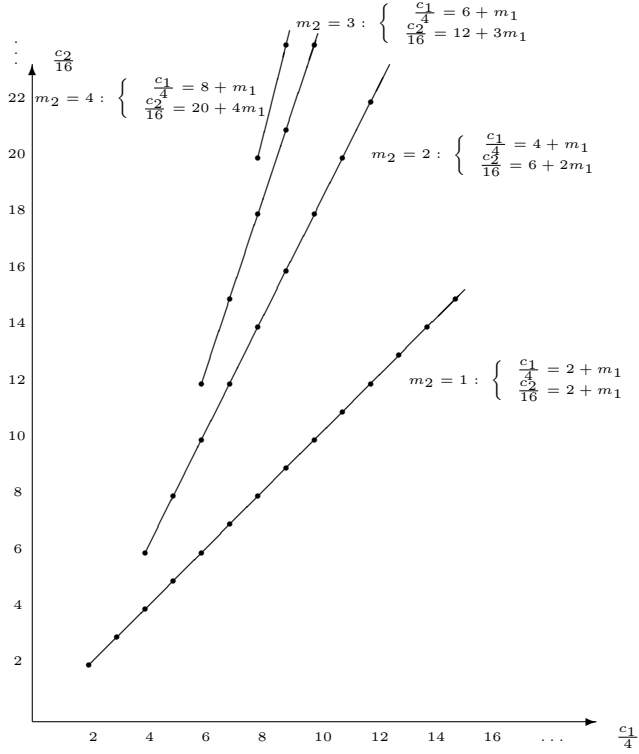
\begin{figure}%[h]
{\tiny
\begin{picture}(300,280)(-100,-20)
\unitlength 0.25mm % = 2.845pt
\linethickness{0.4pt}

\put(0,0){\vector(1,0){300}}
\put(0,0){\vector(0,1){350}}
\put(30,30){\line(1,1){200}}
\put(60,90){\line(1,2){130}}
\put(90,180){\line(1,3){62}}
\put(120,300){\line(1,4){17}}

\put(30,-10){\mbox{2}} %\put(30,0){\line(0,1){200}}
\put(60,-10){\mbox{4}} %\put(60,0){\line(0,1){200}}
\put(90,-10){\mbox{6}} %\put(90,0){\line(0,1){200}}
\put(120,-10){\mbox{8}} %\put(120,0){\line(0,1){200}}
\put(150,-10){\mbox{10}} %\put(150,0){\line(0,1){200}}
\put(180,-10){\mbox{12}} %\put(180,0){\line(0,1){200}}
\put(210,-10){\mbox{14}} %\put(210,0){\line(0,1){200}}
\put(240,-10){\mbox{16}} %\put(240,0){\line(0,1){200}}
\put(270,-10){\mbox{$\ldots$}}
\put(310,-10){\mbox{$\frac{c_1}{4}$}}

\put( -10,30){\mbox{2}} %\put(30,0){\line(0,1){200}}
\put(-10,60){\mbox{4}} %\put(60,0){\line(0,1){200}}
\put(-10,90){\mbox{6}} %\put(90,0){\line(0,1){200}}
\put(-10,120){\mbox{8}} %\put(120,0){\line(0,1){200}}
\put( -13,150){\mbox{10}} %\put(150,0){\line(0,1){200}}
\put(-13,180){\mbox{12}} %\put(180,0){\line(0,1){200}}
\put(-13,210){\mbox{14}} %\put(210,0){\line(0,1){200}}
\put(-13,240){\mbox{16}} %\put(240,0){\line(0,1){200}}
\put(-13,270){\mbox{18}} %\put(210,0){\line(0,1){200}}
\put(-13,300){\mbox{20}} %\put(240,0){\line(0,1){200}}
\put(-13,330){\mbox{22}} %\put(240,0){\line(0,1){200}}
\put(-10,350){\mbox{$\vdots$}}
\put(10,350){\mbox{$\frac{c_2}{16}$}}

\put(200,180)
{\mbox{$m_2=1:\ \left\{\begin{array}{c}  \frac{c_1}{4} = 2+m_1 \\ \frac{c_2}{16} = 2+m_1 \end{array}\right.$}}
\put(180,300)
{\mbox{$m_2=2:\ \left\{\begin{array}{c}  \frac{c_1}{4} = 4+m_1 \\ \frac{c_2}{16} = 6+2m_1 \end{array}\right.$}}
\put(140,370)
{\mbox{$m_2=3:\ \left\{\begin{array}{c}  \frac{c_1}{4} = 6+m_1 \\ \frac{c_2}{16} = 12+3m_1 \end{array}\right.$}}
\put(1,330)
{\mbox{$m_2=4:\ \left\{\begin{array}{c}  \frac{c_1}{4} = 8+m_1 \\ \frac{c_2}{16} = 20+4m_1 \end{array}\right.$}}

\put(30,30){\circle*{3}}
\put(45,45){\circle*{3}}
\put(60,60){\circle*{3}}
\put(75,75){\circle*{3}}
\put(90,90){\circle*{3}}
\put(105,105){\circle*{3}}
\put(120,120){\circle*{3}}
\put(135,135){\circle*{3}}
\put(150,150){\circle*{3}}
\put(165,165){\circle*{3}}
\put(180,180){\circle*{3}}
\put(195,195){\circle*{3}}
\put(210,210){\circle*{3}}
\put(225,225){\circle*{3}}

\put(60,90){\circle*{3}}
\put(75,120){\circle*{3}}
\put(90,150){\circle*{3}}
\put(105,180){\circle*{3}}
\put(120,210){\circle*{3}}
\put(135,240){\circle*{3}}
\put(150,270){\circle*{3}}
\put(165,300){\circle*{3}}
\put(180,330){\circle*{3}}

\put(90,180){\circle*{3}}
\put(105,225){\circle*{3}}
\put(120,270){\circle*{3}}
\put(135,315){\circle*{3}}
\put(150,360){\circle*{3}}

\put(120,300){\circle*{3}}
\put(135,360){\circle*{3}}

\end{picture}
}

\caption{\footnotesize The pattern of allowed values of
$\frac{c_1}{4} = 2m_2+m_1$ and $\frac{c_2}{16} = m_2(m_2+m_1+1)$
for the truncated diagrams ${\bf Q}=[2^{m_2},1^{m_1}]$.
The first four branches with the values $m_2=1,2,3,4$ are shown
from which one should pick up the points with integer $m_1$.}
\label{deg2}
\end{figure}

As we already noted if the number of lines in the Young diagram at even $|Q|$ is $2s$, then $c_j(Q)=W_j^{(s)}$, and if it is $2s+1$, then $c_j(Q)=W_j^{(s+1)}$. At the same time, at odd $|Q|$ it is $c_j(Q)=W_j^{(s+1)}$ in the both cases. However, this is the case only for the diagrams with all lengths of lines larger then one: the unit length lines give rise to zeroes in $W_j^{(s)}$, which makes them equal to those with smaller $s$ due to (\ref{trQ}).
If $s\leq 2$, then from the second line of (\ref{cviaq}) one gets
${\bf q}_2 =\frac{1}{4}c_1(Q) - {\bf q}_1$, see also Fig.\ref{deg2}.
If $s\leq 3$, then ${\bf q}_3 =\frac{c_1(Q)}{4} - {\bf q}_1-{\bf q}_2$
and ${\bf q}_2$ is the non-negative integer root of the quadratic equation,
provided by the third line of (\ref{cviaq}),
\be
{\bf q}_2^2 + {\bf q}_2\left({\bf q}_1  - \frac{c_1(Q)}{4} + 1\right)
+ \left(-\frac{c_2(Q)}{16} + {\bf q}_1({\bf q}_1 - 1))\right)
+ \left(\frac{c_1(Q)}{4} - 3\right)\left(\frac{c_1(Q)}{4}  - {\bf q}_1\right) = 0
\ee
etc.

\section{Borel-like representation of the answers}

Now we explain how to present the answers of this paper in a simpler form using the operator $ \widehat{\cal O}(N)$
mentioned  in the Introduction, i.e. to apply the enhanced Borel transform.
The action of this operator on a function of one variable $z$ is the following way: given a power series
\be
F(z)=\sum_kF_kz^k
\ee
the Borel transformation (parameterized by a parameter $N$) is the power series
\be
 \widehat{\cal O}(N)\cdot F(z)={\cal B}_N\Big[F(z)\Big]_z=\sum_k(N;1)_kF_kz^k
\ee
One may notice that all the correlators in the single Miwa deformed model possess such a form (\ref{FQeven}), (\ref{FQodd}). Hence, all of them can be written the Borel transformation of the finite degree polynomial multiplied with the quadratic exponential. For instance, the results (\ref{13}) can be rewritten using the Borel transform:
\be\label{B1}
F_{[1]}(z,N)=\sum_{k=0}(N+1;1)_{2k}{z^{2k+1}\over (2k)!!}={1\over N}{\cal B}_{N}^{(1)}\left[ze^{z^2\over 2}\right]_z\nn\\
\ee
while (\ref{FQeven}) has the form
\be
F_Q =S_Q\{\delta_{k,2}\}\cdot
\sum_{k=0}^\infty (N;1)_{2k}\frac{z^{2k}}{(2k)!!} \sum_{j=0}^k c_{j}(Q) \frac{(k;-1)_j}{(N;1)_{2j}}
=\nn\\
=S_Q\{\delta_{k,2}\}\cdot{\cal B}_N\left[\sum_{k=0}^\infty \sum_{j=0}c_{j}(Q){(k;-1)_j\over (N;1)_{2j}}\frac{z^{2k}}{(2k)!!}\right]_z=
\boxed{S_Q\{\delta_{k,2}\}\cdot
{\cal B}_N\left[e^{z^2\over 2}\sum_j{\B_j({\bf Q})\over (N;1)_{2j}}\left({z^2\over 2}\right)^j\right]_z}
\ee
and (\ref{FQodd}) has the form
\be
F_Q =S'_Q\{\delta_{k,2}\}\cdot
\sum_{k=0}^\infty (N;1)_{2k+1}\frac{z^{2k+1}}{(2k)!!} \sum_{j=0}^kc_{j}(Q)\frac{(k;-1)_j}{(N;1)_{2j+1}}
=\boxed{S'_Q\{\delta_{k,2}\}\cdot{\cal B}_N\left[ze^{z^2\over 2}\sum_j{\B_j({\bf Q})\over (N;1)_{2j+1}}\left({z^2\over 2}\right)^j\right]_z}
\ee
Thus we represent the average of the Schur function as the Borel transform
of a {\it finite}-degree polynomial in $z$ times $e^{z^2\over 2}$.
In a sense, {\bf the entire perturbation series is summed up into the single and universal exponential},
while the {\bf $Q$-dependence is contained in the elementary polynomial prefactor fully defined by the {\it truncation} map} $Q\longrightarrow {\bf Q}$.

\section{Miwa variable of negative multiplicity}

Let us now consider instead of deformation of the Gaussian measure with one Miwa variable given by ${1\over \displaystyle{\det(1-zX)}}$, the deformation given by $\det(1-zX)$, i.e. the generating function of correlators (\ref{p+c}) with $\c_k=-z^k$. It turns out that in this case that the correlators can be expressed through the same quantities $c_j(Q)$. That is, at even sizes of $|Q|$, the corresponding functions $F_Q^{(-)}$ are
\be
F_Q^{(-)} =S_Q\{\delta_{k,2}\}\cdot
\sum_{k=0}^\infty (-1)^k(N;-1)_{2k}\frac{z^{2k}}{(2k)!!} \sum_{j=0}^k c_{j}(Q^\vee) \frac{(k;-1)_j}{(N;-1)_{2j}}
\ee
and, at odd sizes,
\be
F_Q^{(-)} =S'_Q\{\delta_{k,2}\}\cdot
\sum_{k=0}^\infty (-1)^{k+1}(N;-1)_{2k+1}\frac{z^{2k+1}}{(2k)!!} \sum_{j=0}^kc_{j}(Q^\vee)\frac{(k;-1)_j}{(N;-1)_{2j+1}}
\ee
where $Q^\vee$ denotes the transposed Young diagram.

One can again write these expressions in a compact form using the Borel transform:
\be
F_Q^{(-)} =\boxed{S_Q\{\delta_{k,2}\}\cdot
{\cal B}_{-N}\left[e^{-{z^2\over 2}}\sum_j{c_j(Q^\vee)\over (N;-1)_{2j}}\left(-{z^2\over 2}\right)^j\right]_z}
\ee
for even sizes of $|Q|$,
and
\be
F_Q^{(-)} =\boxed{S'_Q\{\delta_{k,2}\}\cdot
{\cal B}_{-N}\left[ze^{-{z^2\over 2}}\sum_j{c_j(Q^\vee)\over (N;-1)_{2j+1}}\left(-{z^2\over 2}\right)^j\right]_z}
\ee
 for odd sizes. 
 
 Note that this Borel transform ${\cal B}_{-N}$ at integer $N$ cuts off the infinite series in $z$ making a polynomial of it since it inserts the factor ${N!\over (N-k)!}$. On one hand, it justifies the name ``Borel transform" since it drastically makes any series convergent: it just leaves first terms of the series. On the other hand, it is quite expected since any average of $\det(1-zX)$ at finite $N$ is clearly a degree $N$ polynomial of $z$.
 
A natural question arises what transposition of $Q$ means for the truncated diagrams ${\bf Q}$.
 They are not just transposed, but this looks like compensating the intriguing asymmetry of the truncation map,
 e.g.
 \be
 {\bf trunc}\Big\{(2r)^{2s}\Big\} = r^{s+1} \ \ \  \ \ \ {\rm  and} \ \ \ \ \ \
 {\bf trunc}\left\{\Big((2r)^{2s}\Big)^\vee\right\}= {\bf trunc}\Big\{(2s)^{2r}\Big\} = s^{r+1}
 \ee 
 We hope to return to this issue elsewhere.

\section{Conclusion}

To conclude, we reported the first results on the perturbation expansion around the superintegrable
Gaussian potential and found certain non-trivial structures in it.
Our consideration is limited in three respects:

\begin{itemize}
\item We parameterized the perturbation $\{\pi_k\}$ of the potential by the Miwa variable $\pi_k = \pm\sum_{a=1}^m z_a^k$.
\item We restricted it to a single Miwa parameter $z$, which makes our deformation very special.
\item We did not try to sum up the perturbation expansion, i.e. did not study the resurgence problem \cite{Mar}
in this case.
\end{itemize}

Still we discovered interesting structures in this perturbation theory, which can have much
broader applicability: enhanced Borel transform and truncation map.

Further research directions are obvious: one should lift the restrictions,
and one should understand the true meaning of the new structures.
It is a good question whether it would lead to a peculiar notion of superintegrable perturbation
of superintegrable theories and whether the associated resurgence will be able to smoothly
connect different superintegrable points, especially those which have different Stokes indices
(like monomial potentials of different degrees $X^p$ which have $p-1$  Stokes sectors).
Another obvious question is about the $\beta$- and further $(q,t)$-deformations,
which substitute the Schur functions by the Jack and Macdonald polynomials \cite{Mac} and by the Shiraishi functions \cite{Shi,LNS,ELS}.

\section*{Acknowledgements}

This work was supported by the Russian Science Foundation (Grant No.20-12-00195).

\newpage

\section*{Appendix. Details on the Miwa deformed Schur averages
\label{1Miwacors}}

In this Appendix, we provide some evidence in support of the formulas in section \ref{secMiwa1}.

Our original goal was to {\it discover} their structure,
and now it is
to {\it demonstrate} that they are expressed through the polynomials
(\ref{Wpols})
\be
\B_j^{(m)}(a_1\geq a_2 \geq \ldots \geq a_m) :=\ 4^j\cdot\!\!\!\!\!\!\!\!\!\!\!\!\!\!
\sum_{0\leq k_1\leq\ldots\leq k_{m-1}\leq j}\left( \prod_{i=0}^{k_1-1} (a_m-i)  \prod_{i=k_1}^{k_2-1} (a_{m-1}-i) \ldots
\prod_{i = k_{m-1}}^{j-1} (a_1-1)\right)
\label{Wpolsapp}
\ee
The first one is essentially the cutting factorial
\be
W^{(1)}_j(a) = 4^j \cdot \frac{a!}{(a-j)!}
\ee
which is surprisingly similar to what we used to define the {\it enhanced} Borel transform, though its role {\it now} is very different.
For the purposes of this Appendix, it is just a polynomial
\be
W^{(1)}_j(a) =4^j  \cdot \prod_{i=0}^{j-1} (a-i)
\ee
The second one is
\be
W^{(2)}_j(a,b) =   4^j \sum_{k=0}^j \left(\prod_{i=0}^{k-1} (b-i) \prod_{i=k}^{j-1} (a-i)\right)
=4^j\Big(\underbrace{a(a-1)(a-2)\ldots}_{j\ terms}+\underbrace{b(b-2)(b-3)\ldots}_{j\ terms}+abP_j(a,b)\Big)
\label{c2linesprop}
\ee
Its defining property is that it vanishes at all $0\leq b\leq a < j $ complemented with a certain symmetry property.
That is, at $b=0$ from the vanishing condition, one gets $\prod_{i=0}^j(a-i)$,
at $b=1$, it becomes $\prod_{i=1}^j(a-i)$, since $a\geq b=1$, at $b=2$, i.e. for the item $b(b-1)$,
one gets $\prod_{i=2}^j(a-i)$, and so on.
Adding all these items with the same coefficient $\mu_k=1$,
$c_j=4^j \sum_{k=0}^j \mu_k \left(\prod_{i=0}^{k-1} (b-i) \prod_{i=k}^{j-1} (a-i)\right)$,
reflects an additional ``symmetry" relevant for the particular family $[2a+1,2b]$.
Higher polynomial $W_j^{(m)}$ can be deduced from a similar reasoning.

In what follows, we provide tables demonstrating how the correlators (\ref{FQodd1}),
\be\label{FQeven2}
\text{even} \ |Q|: \ \ \ &
F_Q = S_Q\{\delta_{k,2}\}\cdot
\sum_{k=0}^\infty \frac{z^{2k}}{(2k)!!} \sum_{j=0}^k  \frac{k!}{(k-j)!}\frac{(N+2k-1)!}{(N+2j-1)!}\, c_j(Q)
\nn \\
\text{odd} \  |Q|: \ \ \ &
F_Q =S'_Q\{\delta_{k,2}\}\cdot
\sum_{k=0}^\infty \frac{z^{2k+1}}{(2k)!!} \sum_{j=0}^k \frac{k!}{(k-j)!}\frac{(N+2k)!}{(N+2j)!} \, c_j(Q)
\label{FQodd2}
\ee
are expressed through these polynomials $W_j^{(m)}$.
The rule appears to be  (\ref{trunc}),
\be
c_j(Q) = W_j({\bf Q})
\ee
and the content of examples below was summarized in the truncation table
(\ref{trunctable}) in section \ref{secMiwa1}.

\subsection*{Examples of coefficients $c_j(Q)$ for diagrams of given even size}

\be
\begin{array}{c|c|ccccccc|cc}
Q\ \ \backslash\ \  j  & S_Q\{\delta_{k,2}\}& 0 & 1 & 2 & 3 &   & \ldots &&  \\
\hline
\phantom.[11] && 1  &&&&&&&& \B^{(1)}(0)\\
\phantom.[2] &&1 & 4 &&&&&& 4^j\cdot \frac{1!}{(1-j)!}=& \B_j^{(1)}(1)\\
\hline
\phantom.[1111] && 1 &&&&&&&&\B^{(1)}(0) \\
\phantom.[211] &&1 & 8 &&&&&& 8^j\cdot \frac{1!}{(1-j)!}=& \B_j^{(2)}(1,1)\\
\phantom.[22]  &&1& 4 &&&&&& 4^j\cdot \frac{1!}{(1-j)!}=& \B_j^{(1)}(1)\\
\phantom.[31]  && 1 &&&&&&&&\B_j^{(1)}(0)\\
\phantom.[4]  &&1& 8 & 32 &&&&&  4^j\cdot \frac{2!}{(2-j)!} =& \B_j^{(1)}(2)   \\
\hline
\phantom.[111111] && 1 &&&&&&&&\B_j^{(1)}(0) \\
\phantom.[21111]  &&1& 12 &&&&&&  12^j\cdot \frac{1!}{(1-j)!} =& \B_j^{(3)}(1,1,1) \\
\phantom.[2211]  &&1& 4   &&&&&& 4^j\cdot \frac{1!}{(1-j)!} =&  \B_j^{(1)}(1)\\
\phantom.[222]  &&1& 8 &&&&&&  8^j\cdot \frac{1!}{(1-j)!} = &  \B_j^{(2)}(1,1)\\
\phantom.[3111] && 1 &&&&&&&& \B_j^{(1)}(0) \\
\phantom.[321] & 0&  &&&&&&&& 0 \\
\phantom.[33]  &&1& 4 &&&&&& 4^j\cdot \frac{1!}{(1-j)!} =&  \B_j^{(1)}(1)\\
\phantom.[411]  &&1& 12 & 48 &&&&  && \B_j^{(2)}(2,1) \\
\phantom.[42]  &&1& 8 & 32 &&&&&  4^j\cdot \frac{2!}{(2-j)!} =& \B_j^{(1)}(2)  \\
\phantom.[51] && 1  &&&&&&&& \B_j^{(1)}(0) \\
\phantom.[6]  &&1& 12 & 96 & 384 &&&& 4^j\cdot \frac{3!}{(3-j)!} =&  \B_j^{(1)}(0) \\
\hline
\phantom.[11111111] && 1 &&&&&&&&\B_j^{(1)}(0) \\
\phantom.[2111111] &&1& 16 &&&&&& 16^j\cdot \frac{1!}{(1-j)!} =& \B_j^{(4)}(1,1,1,1)\\
\phantom.[221111] &&1& 4 &&&&&& 4^j\cdot \frac{1!}{(1-j)!} =& \B_j^{(1)}(1)\\
\phantom.[22211]  &&1& 12 &&&&&&  12^j\cdot \frac{1!}{(1-j)!} =& \B_j^{(3)}(1,1,1) \\
\phantom.[2222]  && 1 &8 &&&&&& 8^j\cdot \frac{1!}{(1-j)!} =& \B_j^{(2)}(1,1)\\
\phantom.[311111] && 1&&&&&&&& \B_j^{(1)}(0)  \\
\phantom.[32111] & 0 && &&&&&&& 0  \\
\phantom.[3221] && 1  &&&&&&&& \B_j^{(1)}(0)\\
\phantom.[3311] && 1 &4 &&&&&& 4^j\cdot \frac{1!}{(1-j)!} =& \B_j^{(1)}(1)\\
\phantom.[332] && 1 &8 &&&&&&  8^j\cdot \frac{1!}{(1-j)!} =& \B_j^{(2)}(1,1)\\
\phantom.[41111]  &&1& 16 & 64  &&&&& &  \B_j^{(3)}(2,1,1) \\
\phantom.[4211]  &&1& 8 & 32 &&&&&  4^j\cdot \frac{2!}{(2-j)!} =& \B_j^{(1)}(2)\\
\phantom.[422]  &&1& 12 & 48 &&&&&&  \B_j^{(2)}(2,1) \\
\phantom.[431]  &&1& 16 & 96  &&&&&&  \B_j^{(2)}(2,2)  \\
\phantom.[44]  &&1& 8 &32  &&&&&   4^j\cdot \frac{2!}{(2-j)!} =& \B_j^{(1)}(2) \\
\phantom.[5111] && 1 &&&&&&&& \B_j^{(1)}(0)  \\
\phantom.[521] & 0 &&&&&&&&& 0  \\
\phantom.[53]  &&1& 4  &&&&&& 4^j\cdot \frac{1!}{(1-j)!}=& \B_j^{(1)}(1) \\
\phantom.[611] &&1& 16 &   128     & 512  &&&&&   \B_j^{(2)}(3,1)\\
\phantom.[62]  &&1& 12 &  96    &  384  &&&&   4^j \cdot \frac{3!}{(3-j)!} =&  \B_j^{(1)}(3) \\
\phantom.[71] && 1  &&&&&&&& \B_j^{(1)}(0)\\
\phantom.[8]    &&1 & 16 &   192    & 1536        &  6144      &&  & 4^j \cdot \frac{4!}{(4-j)!} =&  \B_j^{(1)}(4)\\
\hline
\ldots
\end{array}
\label{listc}\label{Feven}
\ee

We do not list
the values of $S_Q\{\delta_{k,2}\}$, except for the cases when they vanish.
They are known in general \cite{Pop,MMNO}, see formula (\ref{S2}), and their particular values are
\be
S_{[2a,2b]}\{\delta_{k,2}\}={1\over (2a)!!(2b)!!}\nn\\
S_{[2a+1,2b+1]}\{\delta_{k,2}\}=-{1\over (2a+2)!!(2b)!!}\nn\\
S_{[2a,2b,2c]}\{\delta_{k,2}\}={a+1-c\over a+1}{1\over (2a)!!(2b)!!(2c)!!}\nn
\ee
\be
\ldots
\ee
\be
S_{[2a_1,2a_2,\ldots]}\{\delta_{k,2}\}=\prod_{k=1}\prod_{j=1}{a_k+j-a_{k+2j}\over a_k+j}{1\over\prod_{i=1} (2a_i)!!}\nn\\
S_{[2a_1+1,2a_2+1,2a_3,2a_4,\ldots]}\{\delta_{k,2}\}=\prod_{j=2} {a_1+j-a_{2j}\over a_1+j}\prod_{j=1}{a_2+j-a_{1+2j}
\over a_2+j}\prod_{k=3}\prod_{j=1}{a_k+j-a_{k+2j}\over a_k+j}{1\over (2a_1+2)!!\prod_{i=2}(2a_i)!!}\nn\\
\nn
\ee

The Table implies that
\be
c_j([2a,2b])=4^jj!\binom{a}{j}\nn\\
c_j([2a+1,2b+1])=4^jj!\binom{b}{j}
\label{c2line}
\ee

Thus, we see that the vanishing Gaussian correlators with $S_Q\{\delta_{k,2}\}=0$
remain vanishing under deformation, and this is a result of a non-trivial cancellation in the sum (\ref{FQdef}),
particular terms in it do not vanish themselves.

All other Schur averages are essentially modified by the Miwa deformation,
and become infinite series in $z$ already for a single Miwa variable $z$.
Still there is an additional hierarchy in their complexity labeled by the number of
non-vanishing entries in (\ref{listc}).
The simplest items are the averages with the only non-vanishing entry $c_0(Q)=1$ in the first column.

\subsection*{Examples of coefficients $c_j(Q)$ for diagrams of given odd size}
\be
\begin{array}{c|c|cccccc|ccc}
Q\ \ \backslash\ \  j & S'_Q\{\delta_{k,2}\}  & 0& 1 & 2 & 3 &    & \ldots &&&  \\
\hline
\phantom.[1] & 1  &1 &&&&& &&\B^{(1)}(0)\\
\hline
\phantom.[111] &-\frac{1}{2}  &1 &&&&&&&\B^{(1)}(0)  \\
\phantom.[21] &0  & &&&&& && 0  \\
\phantom.[3] &\frac{1}{2}  & 1 &4 &&&&& 4^j\cdot \frac{1!}{(1-j)!}=& \B^{(1)}(1)  \\
\hline
\phantom.[11111] & \frac{1}{8} &1 &&&&  &&&\B^{(1)}(0) \\
\phantom.[2111] &0 &  &&&&& && 0   \\
\phantom.[221]  &\frac{1}{8}& 1 &&&&& &&\B^{(1)}(0) \\
\phantom.[311]  & -\frac{1}{4} & 1 &  4 &&&& & 4^j\cdot \frac{1!}{(1-j)!} =& \B^{(1)}(1)\\
\phantom.[32]  &\frac{1}{8}& 1 & 8  &&&& &  8^j\cdot \frac{1!}{(1-j)!}=& \B^{(2)}(1,1) \\
\phantom.[41] & 0 &&&& &&&& 0 \\
\phantom.[5] & \frac{1}{8} & 1 &8 &32 &&&& 4^j\cdot \frac{2!}{(2-j)!}= & \B^{(2)}(2,0)\\
\hline
\phantom.[1111111] & -\frac{1}{48} & 1 &&&& &&&\B^{(1)}(0)\\
\phantom.[211111] & 0 &&&&&& && 0\\
\phantom.[22111] & -\frac{1}{24} & 1 &&&& &&&\B^{(1)}(0)\\
\phantom.[2221] & 0 &&&&&& && 0\\
\phantom.[31111] & \frac{1}{16} & 1 &4 &&&& & 4^j\cdot \frac{1!}{(1-j)!}=& \B^{(1)}(1)\\
\phantom.[3211] & -\frac{1}{48} & 1& 12 &&&& &  12^j\cdot \frac{1!}{(1-j)!}=& \B^{(3)}(1,1,1)\\
\phantom.[322] & \frac{1}{16} & 1 &8&&&& & 8^j\cdot \frac{1!}{(1-j)!} =& \B^{(2)}(1,1)\\
\phantom.[331] & -\frac{1}{16} & 1 &4&&&& & 4^j\cdot \frac{1!}{(1-j)!} =& \B^{(1)}(1) \\
\phantom.[4111] & 0 &&&&&& && 0\\
\phantom.[421] & \frac{1}{48} & 1 && &&&&&\B^{(1)}(0) \\
\phantom.[43] & 0 &&&&& &&& 0\\
\phantom.[511] & -\frac{1}{16} & 1 &8&32&&& & 4^j\cdot \frac{2!}{(2-j)!} =& \B^{(2)}(2,0)\\
\phantom.[52] & \frac{1}{24} & 1 &12&48&&&&&     \B^{(2)}(2,1) \\
\phantom.[61]& 0 &&&&&  &&& 0\\
\phantom.[7] & \frac{1}{48} & 1 &12&96&384&& & 4^j\cdot \frac{3!}{(3-j)!}= & \B^{(1)}(3)\\
\hline
\ldots &&&&&&
\end{array}
\label{Fodd}
\ee
This time the vanishing averages, those with $S'_Q\{\delta_{k,2}\}=0$
have nothing to do with the naive properties of Gaussian correlators,
which all vanish for $S_Q$ with $Q$ of odd sizes.
This corresponds to the fact that the series (\ref{FQodd}) begins with $z$ in the first power, and $S'_Q\{\delta_{k,2}\}=0$ is identical for all {\it even} sizes, while, at odd sizes, it vanishes sporadically.

For odd first entry $Q=[q_1,\ldots]$ $j\leq \frac{q_1-1}{2}$.
For even $q_1$ $j\leq 1$, unless $C(Q)=0$ and the whole $F_Q=0$.

\subsection*{Coefficients $c_j(Q)$ for the odd size diagrams with given number of lines}

The structure behind these formulas can be clarified by
the answers for the $2$-line diagrams $Q$.
For half of them $F_Q$ vanish:
\be
F_{[2a,2b-1]}= 0
\ee
Another half is more sophisticated:
\be
\begin{array}{c|c|cccccc|cccc}
Q\ \ \backslash\ \  j & S'_Q\{\delta_{k,2}\}  & 0& 1 & 2 & 3 & 4 &  \ldots& &&  \\
\hline
\phantom.[3,0] & 1/2 & 1&4 &&&&&   4^j\cdot \frac{1!}{(1-j)!} =&  \B_j^{(1)}(1)  =& \B_j^{(2)}(1,0)    \nn \\
\phantom.[3,2] & 1/8 &  1 &8 &&&&&  8^j\cdot \frac{1!}{(1-j)!}  =&& \B_j^{(2)}(1,1)    \nn \\
\hline
\phantom.[5,0]& 1/8 & 1&8&32  &&&&   4^j\cdot \frac{2!}{(2-j)!} = & \B_j^{(1)}(2)=& \B^{(2)}(2,0) \nn\\
\phantom.[5,2]& 1/24 & 1&12&48 & &&&  &&W^{(2)}(2,1)   \nn\\
\phantom.[5,4]& 1/192&  1&16&96 &&&&   && W^{(2)}(2,2)  \nn\\
\hline
\phantom.[7,0]& 1/48 & 1&12&96&384&&  &  4^j\frac{3!}{(3-j)!} =& \B_j^{(1)}(3)=& \B_j^{(2)}(3,0) \nn\\
\phantom.[7,2]& 1/128& 1&16&128&512 &&&  && W_j^{(2)}(3,1)\nn\\
\phantom.[7,4]&1/768& 1&20&192&768 &&&   && W_j^{(2)}(3,2)\nn\\
\phantom.[7,6]&1/9216& 1&24&288&1536 &&&  && W_j^{(2)}(3,3) \nn\\
\hline
\phantom.[9,0]&1/384&1&16&192&1536&6144 &&   4^j\frac{4!}{(4-j)!} =& \B_j^{(1)}(4) =& \B_j^{(2)}(4,0)\nn\\
\phantom.[9,2]&  1/960& 1&20&240&1920&7680& &   && \B_j^{(2)}(4,1)   \nn\\
\phantom.[9,4]&1/5120& 1&24&320&2560&10240 &&   && \B_j^{(2)}(4,2)   \nn\\
\phantom.[9,6]&1/46080 &1& 28& 432& 3840& 15360 &&  && \B_j^{(2)}(4,3) \nn \\
\phantom.[9,8]& 1/737280 &1& 32& 576& 6144& 30720 &&& & \B_j^{(2)}(4,4) \nn\\
\hline
\ldots
\end{array}
\nn \\
\Longrightarrow \ \ \ c_j([2a+1,2b]) = W^{(2)}(a,b)
\ee
where actually
\be
S'_Q\{\delta_{k,2}\}([2a+1,2b])={a+1-b\over (2a+2)!!(2b)!!} \ \ \
\ee

Further,
\be
\begin{array}{c|c|ccccc|cc}
Q\ \ \backslash\ \  j & S'_Q\{\delta_{k,2}\}  & 0& 1 & 2 & 3 & \ldots &  &  \\
\hline
\phantom.[111] & -\frac{1}{2} & 1 &&&&&\delta_{j,0} =\B_j^{(0)}(1)=&\B_j^{(2)}(0,0) \\
\phantom.[311] & -\frac{1}{4} & 1&4&&&& \B_j^{(1)}(1)=&\B_j^{(2)}(1,0)\\
\phantom.[331] & -\frac{1}{16} & 1&4 &&&&  \B_j^{(1)}(1)=&\B_j^{(2)}(1,0) \\
\phantom.[333] & -\frac{1}{64} & 1&8&  &&&&  \B_j^{(2)}(1,1)\\
\phantom.[511] & -\frac{1}{16} & 1&8&32 &&& \B_j^{(1)}(2)=&\B_j^{(2)}(2,0)\\
\phantom.[531] & -\frac{1}{64} & 1&8&32 &&& \B_j^{(1)}(2)=&\B_j^{(2)}(2,0)\\
\phantom.[533] & -\frac{1}{192} & 1&12&48 & &&&  \B_j^{(2)}(2,1)\\
\phantom.[555] & -\frac{1}{9216} & 1 &16 &96    & &&& \B_j^{(2)}(2,2)  \\
\phantom.[711] &-\frac{1}{96} & 1&12&96 &384&& \B_j^{(1)}(3)=& \B_j^{(2)}(3,0)\\
\phantom.[731] &-\frac{1}{384}  & 1&12&96 &384&& \B_j^{(1)}(3)=&\B_j^{(2)}(3,0)\\
\ldots
\end{array}
\nn \\
\ \ \Longrightarrow \ \ c_j([2a+1,2b+1,2c+1]) = \B^{(2)}_j(a,c)
\ee

\be
\begin{array}{c|c|cccc|c}
Q\ \ \backslash\ \  j & S'_Q\{\delta_{k,2}\}  & 0& 1 & 2  & \ldots &  \\
\hline
\phantom.[322] & {1\over 16} & 1 &8&& &\B_j^{(2)}(1,1) \\
\phantom.[522] &  \frac{1}{48} & 1 &12&48& &\B_j^{(2)}(2,1) \\
\phantom.[542] & \frac{1}{384} & 1&16&96 &  &\B_j^{(2)}(2,2)\\
\phantom.[544] & \frac{1}{1536} & 1&16 &96 &   &  \B_j^{(2)}(2,2)  \\
\ldots &&&&&\\
\hline
&&&&&\\
\phantom.[432] & 0 & &&&& 0  \\
\phantom.[632] & 0 & &&&& 0  \\
\phantom.[652] & 0 & &&&& 0  \\
\phantom.[654] & 0 & &&&& 0  \\
\ldots &&&&&\\
\hline
&&&&& \\
\phantom.[221] & {1\over 8} & 1&&  &&  \B_j^{(2)}(0,0)\\
\phantom.[421] & {1\over 48} & 1&  &&&  \B_j^{(2)}(0,0)\\
\phantom.[441] &\frac{1}{96} & 1&4 &&  & \B_j^{(2)}(1,0)\\
\phantom.[443] & \frac{1}{384} & 1&8& &  & \B_j^{(2)}(1,1)\\
\phantom.[621] & \frac{1}{384} & 1& &   &&  \B_j^{(2)}(0,0)\\
\phantom.[641] & \frac{1}{768} & 1 & 4     &&& \B_j^{(2)}(1,0) \\
\phantom.[643] & \frac{1}{3072} & 1&8&& & \B_j^{(2)}(1,1)\\
\phantom.[661] &\frac{1}{3072}  & 1&8&32   && \B_j^{(2)}(2,0)\\
\phantom.[663] &\frac{1}{9216}  & 1&12&48  & &  \B_j^{(2)}(2,1)\\
\phantom.[665] &\frac{1}{73728}  & 1&16& 96   & & \B_j^{(2)}(2,2)\\
\ldots
\end{array}
\nn \\
\ \ \Longrightarrow \ \ \left\{\begin{array}{c} c_j([2a+1,2b,2c]) = \B^{(2)}_j(a,b)   \\
                                                c_j([2a,2b+1,2c]) = 0 \\
                                                c_j([2a,2b,2c+1]) = \B^{(2)}_j(b-1,c)  \end{array}\right.
\ee

\newpage

\subsection*{Coefficients $c_j(Q)$ for the even size diagrams with four lines}

We end with a list of examples for even size diagrams with four lines that illustrate formulas (\ref{mpeven}):

\be
\begin{array}{c|cccccc|c}
Q\ \ \backslash\ \  j  & 0 & 1 & 2 & 3 & 4  & \ldots &  \\
\hline
\phantom{.}[1111] & 1 &&&&&& \B^{(1)}(0) \\
\phantom{.}[3111] & 1 &&&&&& \B^{(1)}(0) \\
\phantom{.}[3311] & 1&4 &&&&& \B^{(1)}(1) \\
\phantom{.}[3331] & 1&4 &&&&& \B^{(1)}(1) \\
\phantom{.}[3333] & 1&8 &&&&& \B^{(2)}(1,1) \\
\phantom{.}[5333] & 1& 8&&&&& \B^{(2)}(1,1) \\
\ldots &&&&&& \\
\hline
\phantom{.}[2211] & 1&4 &&&&& \B^{(1)}(1) \\
\phantom{.}[4211]  & 1&8 &&&&& \B^{(2)}(1,1) \\
\phantom{.}[4411]  & 1&8&32&&&& \B^{(1)}(2) \\
\phantom{.}[3221] & 1 &&&&&& \B^{(1)}(0) \\
\phantom{.}[4321] &  1&8&32&&&& \B^{(1)}(2) \\
\phantom{.}[4431] &  1&8&32&&&& \B^{(1)}(2) \\
\ldots &&&&&& \\
\hline
\phantom{.}[2222] & 1&8 &&&&& \B^{(2)}(1,1) \\
\phantom{.}[4222] & 1&12&48 &&&& \B^{(2)}(2,1) \\
\phantom{.}[4422] & 1&12&48 &&&& \B^{(2)}(2,1) \\
\phantom{.}[4442] & 1&16&96 &&&& \B^{(2)}(2,2) \\
\phantom{.}[4444] & 1 &16&96 &&&& \B^{(2)}(2,2) \\
\phantom{.}[6222] & 1 & 16 & 128 &512&&& \B^{(2)}(3,1) \\
\phantom{.}[8222] & 1 & 20 & 240 & 1920&7680&& \B^{(2)}(4,1)  \\
\phantom{.}[6422] & 1 & 16 & 128 &&&& \B^{(2)}(3,1) \\
\ldots &&&&&&
\end{array}
\ee
We omit the first column with $S_Q\{\delta_{k,2}\}$ in this table, because
they do not vanish in these examples.

\end{document}